\documentclass[12pt,twocolumn]{article}
\usepackage{graphicx}
\usepackage{amsmath}
\usepackage{bm}

\usepackage{booktabs}
\usepackage{graphicx}

\usepackage{adjustbox}
\usepackage{pgfplots} 
\usepackage[table]{colortbl}
\usepackage[utf8]{inputenc}
\usepackage[T1]{fontenc}
\usepackage{authblk}
\usepackage{tabularx}
\usepackage{rotating}
\usepackage[left=1.7cm, right=1.7cm, top=1.7cm, bottom=1.7cm]{geometry}
\usepackage[singlelinecheck=false]{caption}

\usepackage{inputenc}

\usepackage{colortbl}

\usepackage[authoryear,numbers]{natbib}
\setcitestyle{citesep={,},aysep={,}}

\DeclareCaptionFont{small}{\small}
\usepackage{sectsty}
\sectionfont{\fontsize{12}{15}\selectfont}
\subsectionfont{\normalfont\itshape\fontsize{12}{15}\selectfont}
\subsubsectionfont{\normalfont\itshape}

\newcolumntype{Y}{>{\centering\arraybackslash}l}

\usepackage[numbers]{natbib}
\setcitestyle{citesep={,},aysep={,}}

\usepackage[hyphens]{url}

\expandafter\def\expandafter\UrlBreaks\expandafter{\UrlBreaks
      \do\a\do\b\do\c\do\d\do\e\do\f\do\g\do\h\do\i\do\j%
      \do\k\do\l\do\m\do\n\do\o\do\p\do\q\do\r\do\s\do\t%
      \do\u\do\v\do\w\do\x\do\y\do\z\do\A\do\B\do\C\do\D%
      \do\E\do\F\do\G\do\H\do\I\do\J\do\K\do\L\do\M\do\N%
      \do\O\do\P\do\Q\do\R\do\S\do\T\do\U\do\V\do\W\do\X%
      \do\Y\do\Z}

\begin{document}

\title{Accurate and robust segmentation of neuroanatomy in T1-weighted MRI by combining spatial priors with deep convolutional neural networks}

\author{Philip Novosad$^{a,b,*}$, Vladimir Fonov$^{a,b}$, D. Louis Collins$^{a,b}$ and the Alzheimer's Disease Neuroimaging Initiative$^c$}
\date{
\small{
$^a$McConnell Brain Imaging Centre, Montreal Neurological Institute, McGill University, Montreal, Canada \\
$^b$Department of Biomedical Engineering, McGill University, Montreal, Canada \\
$^c$Data used in preparation of this article were obtained from the Alzheimer's Disease Neuroimaging Initiative (ADNI) database (adni.loni.usc.edu). As such, the investigators within the ADNI contributed to the design and implementation of ADNI and/or provided data but did not participate in analysis or writing of this report. A complete listing of ADNI investigators can be found at \url{http://adni.loni.usc.edu/wp-content/uploads/how_to_apply/ADNI_Authorship_List.pdf} \\
$^*$Corresponding author. E-mail address: philip.novosad@mail.mcgill.ca}}

\maketitle

\section{Abstract}
Neuroanatomical segmentation in magnetic resonance imaging (MRI) of the brain is a prerequisite for volume, thickness and shape measurements. This work introduces a new highly accurate and versatile method based on 3D convolutional neural networks for the automatic segmentation of neuroanatomy in T1-weighted MRI. In combination with a deep 3D fully convolutional architecture, efficient linear registration-derived spatial priors are used to incorporate additional spatial context into the network. An aggressive data augmentation scheme using random elastic deformations is also used to regularize the networks, allowing for excellent performance even in cases where only limited labelled training data are available. Applied to hippocampus segmentation in an elderly population (mean Dice coefficient = 92.1\%) and sub-cortical segmentation in a healthy adult population (mean Dice coefficient = 89.5\%), we demonstrate new state-of-the-art accuracies and a high robustness to outliers with the same architecture. Further validation on a multi-structure segmentation task in a scan-rescan dataset demonstrates accuracy (mean Dice coefficient = 86.6\%) similar to the scan-rescan reliability of expert manual segmentations (mean Dice coefficient = 86.9\%), and improved reliability compared to both expert manual segmentations and automated segmentations using FIRST. Furthermore, our method maintains a highly competitive runtime performance (e.g. requiring only 10 seconds for left/right hippocampal segmentation in $1\times1\times1$ mm$^3$ MNI stereotaxic space), orders of magnitude faster than conventional multi-atlas segmentation methods. 

\vspace{2mm}
Keywords: Magnetic resonance imaging, neuroanatomy, segmentation, deep learning, neural networks, spatial priors

\section{Introduction}

Accurate structural segmentation of magnetic resonance (MR) brain images is essential for volume, thickness and shape measurements. Such quantitative measurements are widely used in neuroscience to characterize structural changes associated with age and disease. Given the often prohibitive cost of consistent and reliable expert manual segmentations, a vast number of diverse and fully automated segmentation methods have been proposed. While earlier segmentation methods generally employed various heuristics tailored for the segmentation task at hand, more recent segmentation methods are generally more accurate and attempt to transfer labels from a set of expertly labelled images (atlases) to the target image. Some such methods have attempted to learn complex mappings between image features and labels using traditional machine-learning based classifiers (e.g. support vector machines \citep{Boser1992} and random forests \citep{Breiman2001}) combined with hand-crafted feature sets \citep{Morra2010,Zikic2012}, while others have found success transferring labels using a combination of linear or non-linear image registration with local or non-local label fusion (so-called `multi-atlas segmentation' methods \citep{Coupe2011,Heckemann2006,Iglesias2015}). Indeed, many state-of-the-art results (e.g. hippocampus segmentation \citep{Zandifar2017} and brain extraction \citep{Novosad2018}) exploit a complementary combination of both multi-atlas segmentation and machine-learning methods (e.g. error correction \citep{Wang2011}), though the computational time required for segmentation is usually correspondingly greater.

More recently, convolutional neural networks (CNNs) \citep{Lecun1989} have been used for MR image segmentation, obtaining similar or better performance compared to the previous state-of-the-art while requiring only a fraction of the processing time (despite typically long training times). CNNs are particularly attractive because they have the potential to model much more complicated functions without the need for hand-crafted feature sets, instead autonomously learning to extract task-dependent discriminative features from the training data. Also in contrast to traditional machine-learning classifiers, by stacking many convolutional layers sequentially and/or by incorporating down-sampling operations into the network architectures, CNNs have the capacity to model increasingly complex and long-range spatial relationships in the input, contributing to their excellent performance on image segmentation tasks in particular. However, repeated convolutions and/or down-sampling steps produce coarse features, leading to low-resolution segmentations that can be particularly problematic when targeting smaller structures. Therefore explicitly multi-scale architectures are often preferred, which are capable of preserving local detail while still enabling the modeling of complex long-range spatial relationships. For example, Kamnitsas et al. (\citeyear{Kamnitsas2017}) and Ghafoorian et al. (\citeyear{Ghafoorian2017}) both adopt multi-scale, multi-path architectures which take as input patches extracted at different resolutions, and perform late fusion between the extracted features from the different resolutions. Other works adapt popular architectures such as the U-Net \citep{Ronneberger2015} (adapted in Roy et al. (\citeyear{Roy2018})) and DenseNet \citep{Huang2017} (adapted in Dolz et al. (\citeyear{Dolz2017,Dolz2017b,Dolz2017c})), both of which use skip connections in order to leverage multi-scale information.

Due to hardware limitations of modern graphics processing units (GPUs), modern volumetric medical images (e.g. MR or tomography scans) typically cannot fit into memory, and need to be sub-sampled in order to be processed by a CNN. Most commonly, 3D networks are trained on smaller 3D patches, or 2D networks on single 2D slices. Therefore, despite recent architectures which are capable of better modelling complex, long-range and multi-scale spatial relationships in the input, the implicit spatial context available to the network is still limited. It is therefore often useful to explicitly provide the network with additional spatial contextual information. Examples of applications which leverage spatial contextual features include that of Brebisson et al. (\citeyear{Brebisson2015}), which incorporates distances from pre-defined neuroanatomical structures, that of Wachinger et al. (\citeyear{Wachinger2017}), which incorporates spatial and spectral coordinates (by computing eigenfunctions of the Laplace-Beltrami operator on a pre-estimated brain mask), that of Kushibar et al. (\citeyear{Kushibar2018}), which incorporates non-linear-registration-based atlas probabilities, and that of Ghafoorian et al. (\citeyear{Ghafoorian2017}), which incorporates a hand-crafted combination of such features. While the addition of such features has been shown to result in better performance, the computation of such features is often extremely expensive relative to the time required to apply a trained CNN, limiting the efficiency of the methods.

In this work, we propose a novel CNN-based method for the automated segmentation of neuroanatomy in brain MR images. To maximize spatial context available to the network, we combine a deep 3D fully convolutional neural network with dense connections for efficient multi-scale processing with explicitly provided spatial contextual information through the use of efficient linear-registration-derived spatial priors. We furthermore regularize our trained networks with a data augmentation scheme based on random elastic deformations, increasing the generalizability of the trained networks particularly in cases where limited labelled training subjects are available. In contrast to other recent CNN-based methods for MR segmentation, the scope of this work is also distinguished by the development goals of robustness and versatility. As such, we extensively validate our method on three diverse neuroanatomical segmentation tasks, showing in each case consistently more accurate and robust performance compared to state-of-the-art multi-atlas segmentation and other CNN-based methods, while maintaining a highly competitive runtime performance. Importantly, using a scan-rescan dataset, we also demonstrate that our proposed method achieves an accuracy comparable to the scan-rescan reliability of repeated expert manual segmentations.

\section{Methods and materials}

\subsection{Baseline network} 
In contrast to traditional machine-learning classifiers which treat their inputs as unordered vectors, CNNs explicitly treat their inputs as spatially structured images and work by extracting hierarchical and discriminative representations using sequential applications of the the core building-block known as a `convolutional layer' \citep{Lecun1989}. The function of a convolutional layer is to convolve its input with multiple learned filters and then apply a non-linear activation function (otherwise, the network would just learn a linear transform of the input data). Assuming a simplified network architecture consisting only of convolutional layers, the convolutional filter $W_l^{k,n}$ at network layer $l$ is applied across all the $m_{l-1}$ feature maps produced by the previous convolutional layer $l-1$, resulting in a new set of feature maps, to which a position-wise non-linear activation function $f()$ is applied. For example, the $k$th output feature map at layer $l$ is given by:
\begin{equation}
y_l^k = f(\sum_{n=1}^{m_{l-1}} W_l^{k,n} * x^n_{l-1} + b^k_l)
\end{equation}
where $m_l$ is the number of convolutional filters in layer $l$, $x^n_{l-1}$ is the $n$th feature map of the input to layer $l$, $W_l^{k,n}$ is the $k$th learnable filter, and $b^k_l$ is the learnable bias.
 
We take as our starting point a 3D fully convolutional network, variants of which have shown success in tasks such as brain tumour and ischemic stroke lesion segmentation \citep{Kamnitsas2017}, as well as sub-cortical structure segmentation \citep{Dolz2017}. Instead of using fully connected layers and predicting the label of only one or several voxels for each input patch \citep{Wachinger2017,Ghafoorian2017,Kushibar2018}, fully convolutional networks discard the fully connected layers and produce dense label estimates for whole patches at a time. Consequently, fully convolutional networks have many fewer parameters (and are therefore less prone to overfitting) and preserve the spatial structure of the input. Also following Kamnitsas et al. (\citeyear{Kamnitsas2017}) and Dolz et al. (\citeyear{Dolz2017}), we entirely avoid down-sampling or max-pooling layers to preserve the spatial resolution of the output segmentations.

Our baseline architecture, takes as input a 3D patch with size $25^3$ and $N$ channels (e.g. different MRI contrasts), and returns a smaller 3D patch label estimate with volume $9^3 \times C$ centered on the same respective spatial coordinates in image space, where $C$ is the number of classes. Figure 1 depicts the network architecture schematically, and detailed architectural specifications (including the number of filters and the activation function at each convolutional layer) are provided in Table 1. First, a series of convolutional layers (L$_1$ through L$_8$ in Figure and Table 1) with filters of size $3\times3\times3$ are applied, without padding and with unit stride (in order to preserve spatial resolution). We note that therefore each application of a $3\times3\times3$ convolution layer reduces the size of the input feature maps by 2 voxels in each dimension: after eight $3\times3\times3$ convolutional layers, the size of the feature maps is therefore reduced from $25^3$ to $9^3$. 

\begin{figure*}[!ht]
	\centering
	\includegraphics[scale=0.24]{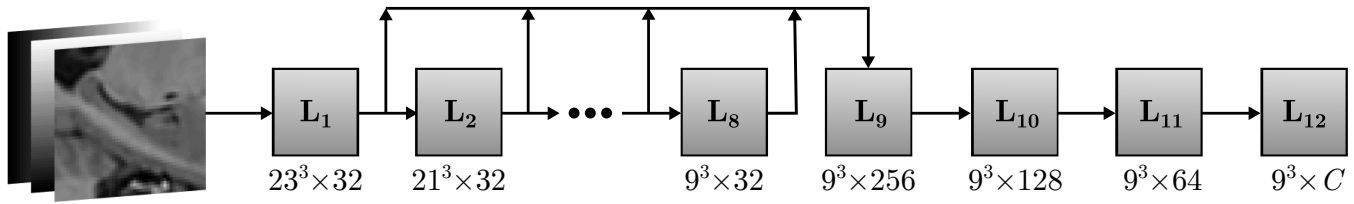} 
    \caption{Schematic of baseline architecture. The network takes as input a $25^3 \times N$ patch (here, spatial coordinates patches are concatenated with the input image patch as described in Section 3.2.1) and returns a multi-channel probabilistic label estimate for the central $9^3$ voxels. The dimensionality of the output of each layer is reported as \textit{size} $\times$ \textit{number of feature maps}. Detailed specifications for each layer are reported in Table 1.}
\label{fig:arch}
\end{figure*} 

\begin{table*}[!ht]
\centering
\fontsize{9}{10}\selectfont
\begin{tabular*}{\linewidth}{l @{\extracolsep{\fill}} llllll }
\hline
 & Operation & Filters & Non-linearity & Input dimension & Output dimension & Notes \\ 
\hline
L$_1$ & Convolution & ($3 \times 3 \times 3) \times 32$ & ELU & $25 \times 25 \times 25 \times N$ & $23 \times 23 \times 23 \times 32$ & --- \\
L$_2$ - L$_8$ & Convolution & ($3 \times 3 \times 3) \times 32$ & ELU & $23 \times 23 \times 23 \times 32$ & $9 \times 9 \times 9 \times 32$ & --- \\
L$_9$ & Dense connection & --- & --- & $(9 \times 9 \times 9 \times 32) \times 8$ & $9 \times 9 \times 9 \times 256$ & BN \\
L$_{10}$ & Convolution & ($1 \times 1 \times 1) \times 128$ & ELU & $9 \times 9 \times 9 \times 256$ & $9 \times 9 \times 9 \times 128$ & DO\\
L$_{11}$ & Convolution & ($1 \times 1 \times 1) \times 64$ & ELU & $9 \times 9 \times 9 \times 128$ & $9 \times 9 \times 9 \times 64$ & DO \\
L$_{12}$ & Convolution & ($1 \times 1 \times 1) \times C$ & Spatial softmax & $9 \times 9 \times 9 \times 64$ & $9 \times 9 \times 9 \times C$& --- \\
\hline
\end{tabular*}
\caption{Baseline network architecture specifications. The network has roughly 220000 parameters (depending on the number of classes $C$ and the number of input channels $N$). For each layer, if applicable, the size and number of learnable filters is reported in the third column as (\textit{filter size}) $\times$ \textit{number of filters}. A corresponding schematic depiction of the network architecture is shown in Figure 1. BN: batch normalization, DO: dropout (with dropout probability 0.1).}
\label{tab:architecture}
\end{table*} 

While the first layers of a CNN extract high-resolution feature maps which respond to basic local image features such as edges, due to repeated convolution, the feature maps extracted from the deeper layers tend to have lower resolution and respond to more global and abstract image features. Ideally, a classifier should consider features extracted across all scales of the input, i.e. features extracted from each convolutional layer rather than only the last. To this end, similar to Dolz et al. (\citeyear{Dolz2017}), we follow Huang et al. (\citeyear{Huang2017}) and use a `dense connection' after layer $L_8$ which consists of the channel-wise concatenation of the feature maps produced by the preceding convolutional layers $L_1$ through $L_8$. In this way, the last convolutional layers following the dense connection have direct access to the multi-scale feature maps produced by each the preceding convolutional layers, and are therefore capable of maintaining feature maps with high spatial resolution while also considering complex and long-range characteristics of the input. The dense connection also encourages feature re-use and improves the convergence properties of the network during training by providing a more direct path during backpropagation between the calculated loss and the earlier convolutional layers \citep{Huang2017}. We note that since the output feature maps produced by each convolutional layers have different sizes, only the central $9^3$ voxels of each feature map are concatenated. The concatenated feature maps are then batch normalized \citep{Ioffe2015} to make the first and second order statistics of feature maps consistent across channels, improving convergence.

The batch-normalized concatenated feature maps are then further processed by two convolutional layers (L$_{10}$ and L$_{11}$ in Figure 1 and Table 1) with filters of size $1\times1\times1$. These layers serve to model inter-channel (and therefore also multi-scale) dependencies and also to reduce the number of feature maps prior to being fed into final classification layer. Dropout \citep{Srivastava2014} (with drop probability $p=0.1$) is applied after both L$_{10}$ and L$_{11}$ to help regularize the model. The final classification layer (L$_{12}$ in Figure 1 and Table 1) processes the resulting features using a set of $C$ filters (where $C$ is the number of classes under consideration) of size $1\times1\times1$, producing a probabilistic label estimate image $p_c$ of size $9 \times 9 \times 9$ for each class $c$. 

Network parameters (i.e. convolutional filters and biases) are estimated iteratively by optimizing the loss function in equation (2) using a gradient-descent optimizer over mini-batches of size $B$. The loss function $\mathcal{L}$ to be minimized is defined as:
\begin{equation}
\mathcal{L} = J + \alpha \| W \|_2^2
\end{equation}
where $\alpha$ (empirically set to $1\times 10^{-4}$ in our experiments) penalizes the $l_2$ norm of the network filters $W$, reducing overfitting, and $J$ is the categorical cross-entropy loss:
\begin{equation}
J = -\frac{1}{B \times V} \sum_{c=1}^C \sum_{b=1}^B \sum_{v=1}^V c^v_b \log p_{c^v_b}
\end{equation}
where $p_{c^v_b}$ is the output of the final classification layer for voxel $v$ and class $c$, and $c^v_b$ is the corresponding reference label. The training procedure is further detailed in Section 3.4.

\subsection{Adding spatial priors}

\subsubsection{Spatial coordinates}
Though other works have explored the effects of explicitly augmenting their architectures with spatial coordinates and/or spatial probability maps, this is typically accomplished by concatenating a single vector to the output of a flattened fully connected layer \citep{Wachinger2017,Ghafoorian2017,Kushibar2018} which has no analogue in the proposed network. Furthermore, as our fully convolutional network makes predictions for whole patches rather than one or several voxels, the features associated with each voxel should be augmented with their respective spatial coordinates rather than that of the central voxel only. We therefore make use of whole spatial coordinate patches: given an input patch centered on spatial coordinate $(x,y,z$) in image space, we extract three additional patches (with the same spatial dimensions as the input patch) centered on spatial coordinate $(x,y,z)$ from each of three `coordinate images'. For example, for the $x$-coordinate image, the value at spatial coordinate ($x,y,z$) is simply $x$, and similar for the $y$- and $z$-coordinate images. The spatial coordinate patches are then concatenated with the image intensity patch before being fed into the first layer of the network.

\subsubsection{Working volumes}
\begin{figure}[!ht]
	\centering
	\includegraphics[scale=0.8]{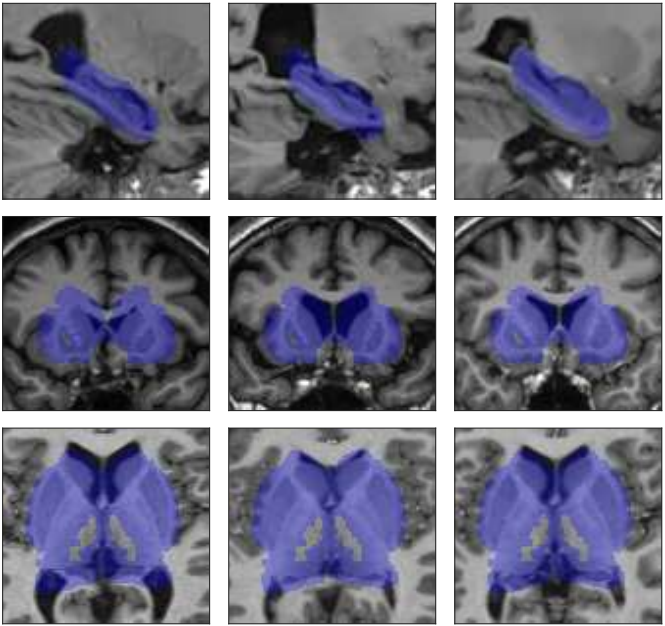} 
    \caption{Example working volumes overlaid on random subjects from the hippocampus (top row),  sub-cortical (middle row) and multi-structure (bottom row) segmentation experiments.}
    \label{fig:sampling_masks}
\end{figure}

In order to benefit from the explicit incorporation of spatial coordinates into the network input, it is important that all images are spatially aligned. In this work we accomplish this by using a light pre-processing pipeline which incorporates linear registration to a common space. For anatomically well defined structures which present relatively little variation in shape and location after registration, we can further take advantage of this registration step by defining, given a set of training subjects, a working volume in which patches are extracted when training and applying the networks (we note that during training, we sample an equal number of patches from each class as described in Section 3.4). For each structure of interest $c$, we first obtain a class-specific boundary-like working volume $B^c$ by subtracting the union $U^c = \bigcup_{i=1}^{I} M_i^c$ from the intersection $I^c = \bigcap_{i=1}^{I} M_i^c$ of all training subject labels $M_i^c$ for the given structure of interest, and then dilating the result, i.e. $B^c = D \oplus (U^c - I^c)$ where $D$ is the dilation structuring element (here set to $3 \times 3 \times 3$ voxels). We also define a corresponding positive volume $P^c = U^c - (U^c \bigcap B^c$). We then define obtain the final working volume $B = \bigcup_{c=1}^{C} B^c$ and the final positive volume $P = \bigcup_{c=1}^{C} P^c$. Example working volumes are shown in Figure \ref{fig:sampling_masks}.

When training the networks, all patches are drawn from the boundary-like working volume $B$. When applying a trained network, the final label estimate is obtained by adding the label estimate within the working volume $B$ with the positive mask $P$. Using such working volumes has several advantages. First, it forces the network to learn from challenging examples, which usually occur near the boundary of the object of interest. Second, it substantially decreases the time required to apply trained networks by removing areas which are highly confidently foreground or background from consideration.

\subsection{Data augmentation with random elastic deformations}

Deep neural networks, which have a high modelling capacity, are particularly dependent on the availability of large quantities of labelled training data in order to generalize well to new unseen test data. In the context of MRI segmentation, low numbers of training samples are typically encountered due to the high cost of generating manually annotated data. To remedy this problem, data augmentation can be used to artificially expand the training set. Commonly, this is accomplished by applying user-specified but label-preserving transformations to the training data, such as reflections, rotations and flips. However, since in the present work all images are linearly registered in a common space and are therefore approximately the same size and with the same orientation, these transformations would be counterproductive. Rather, the relevant differences between linearly registered images are local and non-linear in nature. To create plausible synthetic training samples, we therefore chose to apply random 3D elastic deformations using a method based on Simard et al. (\citeyear{Simard2003}).

To generate a random elastic deformation, we first generate a 3D vector field (where each vector element specifies the per-pixel displacement in each of the $x$, $y$ and $z$ directions respectively) with the same spatial dimensions as the input samples, and then assign each vector element a random value selected from the uniform distribution $U(-1,1)$. The vector field is then smoothed in each direction using Gaussian kernels with standard deviation $\sigma_e$ (controlling the elasticity of the deformation), normalized to have a mean per-pixel displacement of one, and then multiplied by a constant $\alpha_i$ (controlling the intensity of the deformation), producing the final deformation. During training, one such unique random elastic deformation is generated and used to interpolate each training sample (i.e. the image appearance patch, the three spatial coordinates patches, and the reference label image) using linear interpolation. We note that applying linear interpolation introduces a slight blurring in the label images, which itself can be useful as a regularization technique \citep{Szegedy2016}. The parameters $\sigma_e$ and $\alpha_i$ were determined using a coarse grid search, detailed in Section 4.1.3.

\subsection{Training and testing}
Network parameters $\theta$ are optimized iteratively using RMSProp \citep{Tieleman2012}, an adaptive stochastic gradient descent algorithm with Nesterov momentum \citep{Nesterov1983} (momentum = 0.9) for acceleration. At each training iteration, we sample approximately 2,000 voxels, with an equal number of voxels sampled from each training subject. Since CNNs are sensitive to class imbalance, we sample an equal number of voxels from each structure (background included). Training samples (i.e. whole patches) are then extracted around each selected voxel, and image appearance patches are individually normalized to zero mean and unit standard deviation. All training samples are then randomly shuffled and processed by the network in batches of size $B$. Network weights are randomly initialized with the Glorot method \citep{Glorot2010}, and all biases are initialized to zero. Training was performed on a single NVIDIA TITAN X with 12GB GPU memory. Software was coded in Python, and used Lasagne \citep{Lasagne}, a lightweight library to build and train the neural networks in Theano \citep{Theano}. 

To counter over-fitting we employ the early-stopping technique, whereby a randomly selected validation subject set (taken here to be 20\%) is held out from the training subject set. Before training, a fixed validation set is obtained by randomly sampling a fixed number of patches from each validation subject within the working volume. Unlike during the training phase, we extract the validation set uniformly (i.e. without enforcing class balance), so that the distribution of classes in the validation set better approximates the true distribution of classes within the working volume.  During training, at each iteration, the average categorical cross-entropy loss (Equation (3)) over the validation set is measured. The final weights for the trained model are taken from the iteration which achieved the lowest validation loss, and training is stopped if the previously attained lowest validation error does not further decrease after 30 iterations. A static learning rate of $2.5 \times 10^{-4}$ is used for training the networks. This value was empirically determined in our preliminary experiments following the suggestions described by Bengio et al. (\citeyear{Bengio2012}), i.e. by roughly finding the smallest learning rate which caused training to diverge, and then dividing it in half. When training the baseline network, we process the samples in smaller batches of size $B = 128$. 

At testing, we apply the trained network with a stride of 4 voxels in each dimension to  estimate the labels within the working volume only. The label estimate within the working volume is then added to the positive mask (see Section 3.2.2), obtaining the final label estimate. We further apply a fast and simple post-processing step which consists in only keeping the largest connected component for each label, thereby eliminating isolated clusters of false positives.

\section{Experiments and results}
We first assessed the impact of spatial priors, architecture depth and width, and data augmentation on the task of hippocampal segmentation in the ADNI (http://adni.loni.usc.edu) \citep{Mueller2005} dataset. To demonstrate the versatility of the proposed segmentation method, we further applied it sub-cortical segmentation using the IBSR (http://www.nitrc.org/projects/ibsr/) dataset, and multi-structure segmentation using the OASIS (https://www.oasis-brains.org/) \citep{Marcus2007} scan-rescan dataset. Dataset and pre-processing specifications are provided in the respective sections below.

We assess segmentation accuracy and reliability using the Dice coefficient. The Dice coefficient measures the extent of spatial overlap between two binary images:
\begin{equation}
\text{Dice} = 100\% \times 2 {|A \cap R|} / (|A| + |R|)
\end{equation}
where $A$ is an automatically segmented label image, $R$ is the reference label image, $\cap$ is the intersection, and $|\cdot|$ counts the number of non-zero elements. We here express the Dice coefficient as a percentage, with 100\% indicating perfect overlap. We note that for multi-label images, we compute the Dice coefficient for each structure independently. 

Segmentations with high general overlap may nonetheless have clinically important differences in their boundaries. To measure these differences, we also use the modified Hausdorff distance (MHD) \citep{Dubuisson1994}:
\begin{equation}
\text{MHD} = \max(h(A,R),h(R,A)) 
\end{equation}
where $h(A,R)$ is the mean distance of the set of minimum distances between each labelled voxel in $A$ and its nearest labelled voxel in $R$; $h(R,A)$ is computed similarly. Note that for the MHD, lower values are better.

Finally, we assess the statistical significance of differences between distributions of Dice coefficients or MHD values using non-parametric Wilcoxon signed-ranked tests.

\subsection{Application to hippocampus segmentation: effect of spatial priors, architecture and data augmentation}
The hippocampal dataset consists of sixty T1-weighted (T1w) 1.5T scans with manually segmented \citep{Pruessner2000} left and right hippocampi. Twenty subjects were selected from each of the following clinical subgroups: normal controls, mild cognitive impairment, and Alzheimer's disease. Since this dataset was previously used to compare several state-of-the-art algorithms \citep{Zandifar2017}, for our experiments, we use the same data (e.g. previously pre-processed as described in Zandifar et al. to enable meaningful comparisons with the results reported in the aforementioned work. Pre-processing consisted of patch-based denoising \citep{Coupe2008}, N3 non-uniformity correction \citep{Sled1998}, linear intensity normalization, and affine registration to the MNI-ICBM152 template \citep{Fonov2011} with $1 \times 1 \times 1$ mm$^3$ resolution. We trained our networks to segment both the right and left hippocampi. To obtain a segmentation for each subject, we carried out a 5-fold cross-validation. 

\subsubsection{Effect of spatial priors}
Mean Dice and MHD values for several variants of the baseline network (CNN-B) are reported in Table 2. We also include in Table 2 results without the post-processing step (keeping only the largest connected component for each segmented structure, i.e. removing isolated clusters of false positives). Post-processing was crucial for obtaining good performance with CNN-B, (e.g. reducing the mean MHD from 4.59 mm to 0.27 mm), but was less important when applied to the methods using either working volumes or spatial coordinates, and still less important when applied to the method incorporating both spatial priors (CNN-SP). For the fairest possible comparison, we below compare the different methods when combined with post-processing.

Augmenting CNN-B with spatial coordinates (CNN-SC) only, or performing the training and testing within the working volume (CNN-WV) only, both resulted in statistically significant ($p \leq 10^{-3}$) increases in performance with respect to both mean Dice and mean MHD. Combining both spatial priors resulted in the best performance (mean Dice = 91.5\%, mean MHD = 0.23 mm), a statistically significant improvement over both CNN-SC and CNN-WV with respect to both mean Dice ($p \leq 0.005$) and mean MHD ($p \leq 0.01$). Using the working volume also drastically increased computational efficiency reducing the mean processing time from 28.3 $\pm$ 1.7 seconds per subject to 3.5 $\pm$ 0.5 seconds per subject. 

\begin{figure*}[!ht]
    \centering
    \includegraphics[scale=0.8]{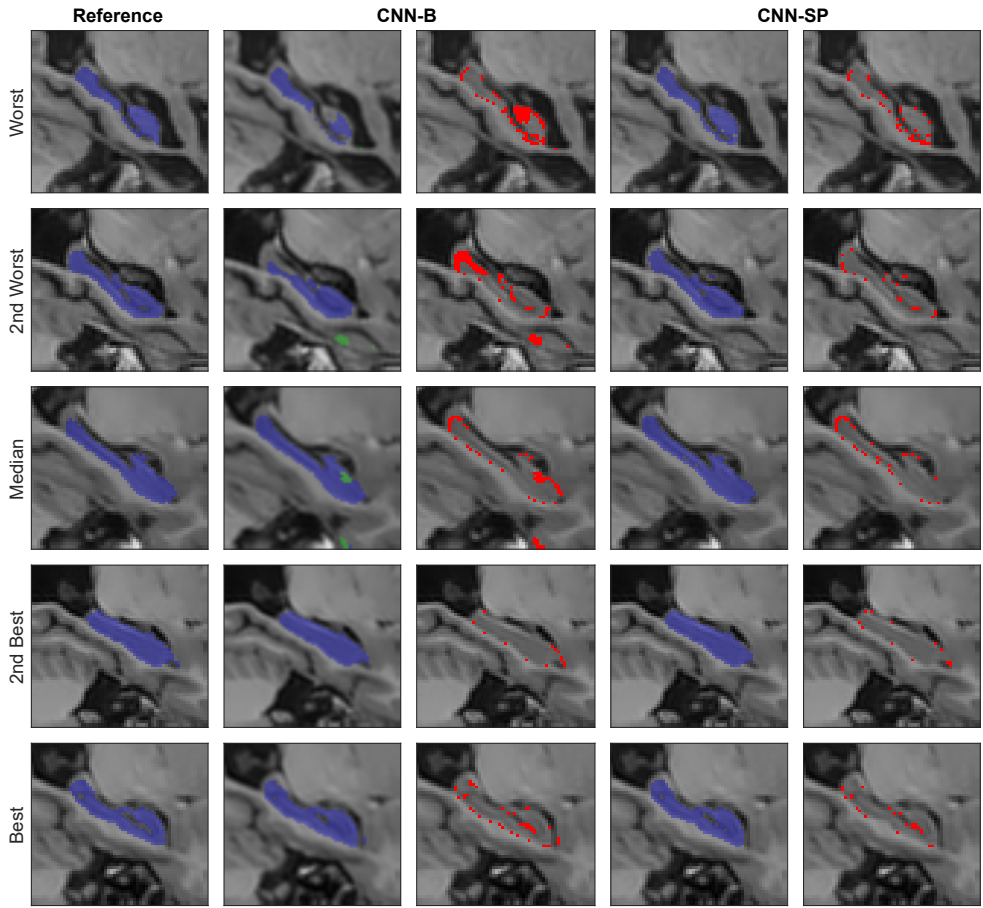}
    \caption{Example right hippocampus segmentations and respective errors using the baseline network (CNN-B), and the network augmented with both the working volume and spatial coordinates (CNN-SP). The subjects with the worst, second worst, median, second best and best overlaps after applying CNN-B are shown for comparison. We note that without the spatial priors, the network has more difficulty distinguishing between the left (overlaid in green) and right (overlaid in blue) hippocampi, and produces isolated clusters of false positives. Errors are overlaid in red in columns three and five.}
    \label{fig:ex_seg_sc_wv}
\end{figure*}

Combining both spatial priors (CNN-SP) resulted in the best performance (mean Dice = 91.5\%, mean MHD = 0.23 mm), a statistically significant improvement over both CNN-SC and CNN-WV with respect to both mean Dice ($p \leq 0.005$) and mean MHD ($p \leq 0.01$). Example segmentations showing the improvements due to the use of spatial priors are displayed in Figure \ref{fig:ex_seg_sc_wv}. Note that these segmentations are shown without post-processing (consisting of keeping only the largest connected component for each label) to better understand the origin of the errors made by the two methods. One can see that without spatial priors, the baseline network CNN-B has more difficulty distinguishing between left and right hippocampi, and produces isolated clusters of false positives. Both of these errors, as shown in Figure \ref{fig:ex_seg_sc_wv}, can be largely addressed by supplying the network with adequate spatial context.

\begin{table}[!ht]
\centering
\fontsize{9}{10}\selectfont
\begin{tabular*}{\linewidth}{l @{\extracolsep{\fill}} lll }
\hline
 & Dice (\%) & MHD (mm) \\ 
\hline
CNN-B & 90.2 (2.8) & 0.27 (0.10) \\
CNN-B$^\dagger$ & 87.2 (3.2) & 4.59 (2.47) \\
CNN-WV & 90.9 (2.4) & 0.26 (0.13)  \\
CNN-WV$^\dagger$ & 90.6 (2.5) & 0.54 (0.59) \\
CNN-SC & 91.3 (2.1) & 0.24 (0.09)  \\
CNN-SC$^\dagger$ & 90.1 (2.6) & 0.64 (0.40) \\
CNN-SP & 91.5 (2.0) & 0.23 (0.08)   \\
CNN-SP$^\dagger$ & 91.4 (2.1) & 0.25 (0.14) \\
\hline
\end{tabular*}
\caption{Effect of augmenting the baseline network (CNN-B) with spatial coordinates alone (CNN-SC), the working volume alone (CNN-WV), and both spatial priors (CNN-SP) on network performance for the hippocampus segmentation experiment. The $\dagger$ superscript indicates that no post-processing was performed. Mean Dice and MHD values over both left and right hippocampi are reported, with standard deviations in parentheses.}
\end{table}  

\subsubsection{Architectural modifications}
We assessed whether the performance of the CNN-SP method could be further improved by widening (learning more filters per convolutional layer) or deepening (including more convolutional layers) the network architecture. We note that for the deeper networks, we correspondingly increased the size of the input samples in order to preserve the output size. Also, because of the increased memory requirements associated with deeper architecture, during training, we reduced the batch size to $B = 32$ as needed. Quantitative results are reported in Table 3. Widening the network by doubling the number of learnable filters (from 32 to 64) in convolutional layers L$_1$ through L$_8$ produced no appreciable gain in performance. However, deepening the network by increasing the number of 3$\times$3$\times$3 convolutional layers resulted in a gradual increase in performance with respect to both mean Dice and MHD, with a plateau reached when using eighteen or twenty such convolutional layers (corresponding to input samples with spatial dimensions $39^3$ or $41^3$ respectively). We note that the mean runtime of the deepest networks was correspondingly higher (e.g. ($10.4 \pm 0.5$) seconds per subject with twenty 3$\times$3$\times$3 convolutional layers) compared to CNN-SP (($3.5 \pm 0.5$) seconds per subject). For subsequent experiments, we opt to evaluate the deepest network, and denote the architecture by `CNN-SP-D'.

\begin{table}[!ht]
\centering
\fontsize{9}{10}\selectfont
\begin{tabular*}{\linewidth}{l @{\extracolsep{\fill}} lll }
\hline
 & Dice (\%) & MHD (mm) \\ 
\hline
CNN-SP(12-32) & 91.5 (2.0) & 0.23 (0.08) \\
CNN-SP(12-64) & 91.4 (2.3) & 0.24 (0.12) \\
CNN-SP(14-32) & 91.6 (2.0) & 0.23 (0.07) \\
CNN-SP(16-32) & 91.7 (2.1) & 0.23 (0.08) \\
CNN-SP(18-32) & 92.0 (1.8) & 0.22 (0.06) \\
CNN-SP(20-32) & 91.9 (1.9) & 0.21 (0.07) \\
\hline
\end{tabular*}
\caption{Effect of widening and deepening the CNN-SP architecture on performance for the hippocampus segmentation experiment. The number of $3\times3\times3$ convolutional layers and their associated number of learnable filters are specified in parentheses (e.g. (12-32) specifies 12 convolutional layers each with 32 filters). Mean Dice and MHD values over both left and right hippocampi are reported, with standard deviations in parentheses.}
\end{table}

\subsubsection{Data augmentation}
One concern when training deep convolutional neural networks with high modelling capacities is their increased tendency to overfit the training data, thus generalizing poorly when applied to new unseen testing data. As discussed in Section 3.3, data augmentation can be used to synthesize new training data to increase the generalizability of the trained networks. Using the CNN-SP-D architecture (i.e. twenty $3\times3\times3$ convolutional layers, corresponding to input samples with spatial dimension $41^3$), we assessed the impact of random elastic deformation for data augmentation in two scenarios: the first, in which all available training subjects are used, and the second, in which only a randomly selected subset (25\%, or 12 subjects) of the training subjects in each training fold is used (leaving the test folds unchanged). To determine the two parameters $\sigma_e$ and $\alpha_i$ associated with our data augmentation scheme (see Section 3.3), we conducted a coarse grid search (applied in the second scenario) over $\sigma_e = \{4,8,16\}$ mm and $\alpha_i = \{1,2,4,8\}$ mm, and found the best performance with $\sigma_e = 4$ mm and $\alpha_i = 2$ mm. These parameters are used for data augmentation in the remaining experiments. Using these parameters, example randomly deformed training samples are shown in Figure 4. 

\begin{figure}[!ht]
	\centering
	\includegraphics[scale=0.7]{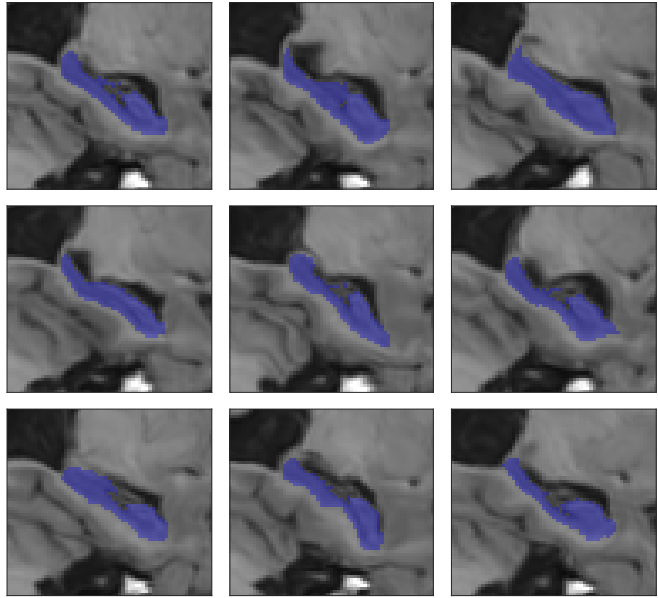} 
    \caption{Example random deformations applied to an original sample (top left) using the parameters $\sigma_c = 4$ mm and $\alpha_i = 2$ mm.}
\end{figure} 

Mean Dice and MHD values are reported in Table 4. As expected, the benefit of using data augmentation was largest when fewer training subjects were used: when using only 25\% of the available training subjects (12 training subjects per fold), training the networks with data augmentation increased the mean Dice coefficient by 1.1\% ($p < 1 \times 10^{-11}$) and reduced the standard deviation by 0.8\% compared to training the networks without augmentation. The relative increase in mean Dice coefficient was reduced to only 0.2\% ($p = 0.06$) when using all available training subjects. With regards to mean MHD, data augmentation resulted in improved performance only in the low-data regime, reducing mean MHD from 0.28 mm to 0.24 mm ($p < 1 \times 10^{-8} $).  

\begin{table}[!ht]
\centering
\fontsize{9}{10}\selectfont
\begin{tabular*}{\linewidth}{l @{\extracolsep{\fill}} lll }
\hline
 & & Dice (\%) & MHD (mm) \\ 
\hline
25\% & CNN-SP-D & 90.0 (2.9) & 0.28 (0.10)   \\
& CNN-SP-D + DA & 91.1 (2.1) & 0.24 (0.08) \\
100\% & CNN-SP-D & 91.9 (1.9) & 0.21 (0.07)   \\
& CNN-SP-D + DA & 92.1 (1.9) & 0.21 (0.07) \\
\hline
\end{tabular*}
\caption{Effect of data augmentation (DA) with random elastic deformations on CNN-SP-D when using either 25\% (12 training subjects per cross-validation fold) or 100\% (48 training subjects per cross-validation subjects) of the available training data for the hippocampus segmentation experiment. Mean Dice and MHD values over both left and right hippocampi are reported, with standard deviations in parentheses.}
\end{table}

\subsubsection{Comparison to other methods}
We further compared several variants of our CNN-based method with several other popular and/or state-of-the-art segmentation methods on the same dataset using the segmentations previously produced in the work of Zandifar et al. (\citeyear{Zandifar2017}) (see Table \ref{tab:hc_comparison}), which includes results for four different methods, both before and after applying error correction (EC) \citep{Wang2011}, a machine learning based wrapper which attempts to correct systematic errors made by the initial host segmentation method. The methods included are FreeSurfer \citep{Fischl2012}, ANIMAL \citep{Collins2010} (a multi-atlas technique combining non-linear registration with majority-vote label fusion), traditional patch-based (PB) segmentation \citep{Coupe2011} (a multi-atlas technique combining linear registration with patch-based label fusion), and an augmented approach combining patch-based segmentation with non-linear registration. 

\begin{table}[!ht]
\centering
\fontsize{8}{10}\selectfont
\begin{tabular*}{\linewidth}{l @{\extracolsep{\fill}} lll }
\hline
 & Left & Right & Both \\
\hline
CNN-B & 90.7 (2.3) & 89.8 (3.2) & 90.2 (2.8) \\
 & 0.25 (0.07) & 0.29 (0.12) & 0.27 (0.10) \\
CNN-SP & 91.5 (1.9) & 91.6 (2.1) & 91.5 (2.0) \\
 & 0.23 (0.07) & 0.23 (0.09) & 0.23 (0.08) \\
CNN-SP-D & \textbf{92.0 (1.6)} & \textbf{91.8 (2.2)} & \textbf{91.9 (1.9)} \\
 & \textbf{0.21 (0.06)} & \textbf{0.22 (0.08)} & \textbf{0.21 (0.07)} \\
CNN-SP-D + DA & \textbf{92.0 (2.0)} & \textbf{92.2 (2.1)} & \textbf{92.1 (1.9)} \\ 
 & \textbf{0.21 (0.07)} & \textbf{0.22 (0.07)} & \textbf{0.21 (0.07)} \\
FreeSurfer & 75.8 (4.7) & 75.6 (4.8) & 75.7 (4.7) \\
 & 0.94 (0.27) & 0.98 (0.24) & 0.96 (0.26) \\
FreeSurfer + EC & 85.9 (3.3) & 86.3 (3.1) & 86.1 (3.2) \\
 & 0.44 (0.13) & 0.42 (0.09) & 0.43 (0.11) \\
ANIMAL & 86.3 (2.6) & 85.9 (3.0) & 86.1 (2.8) \\
 & 0.40 (0.07) & 0.42 (0.09) & 0.41 (0.08) \\
ANIMAL + EC & 86.5 (2.4) & 86.2 (3.0) & 86.4 (2.7) \\
 & 0.43 (0.08) & 0.44 (0.09) & 0.44 (0.08) \\
PBS & 87.5 (2.5) & 87.3 (3.6) & 87.4 (3.1) \\
 & 0.40 (0.09) & 0.40 (0.13) & 0.40 (0.11) \\
PBS + EC & 88.2 (2.5) & 88.2 (3.6) & 88.2 (3.1) \\
 & 0.39 (0.08) & 0.40 (0.14) & 0.39 (0.11) \\
PBS + NLR & 88.3 (2.2) & 88.0 (3.2) & 88.1 (2.7) \\
 & 0.39 (0.07) & 0.39 (0.12) & 0.39 (0.10) \\
PBS + NLR + EC & 89.1 (2.6) & 88.9 (3.1)  & 89.0 (2.6) \\
 & 0.37 (0.06) & 0.37 (0.12) & 0.37 (0.10) \\
\hline
\end{tabular*}
\caption{Comparison of four of our CNN-based segmentation methods with previously reported results \citep{Zandifar2017} for the segmentation of the left and right hippocampi in the ADNI-1 dataset. Each table cell reports the mean Dice coefficient (standard deviation) as a percentage on top and the mean MHD (standard deviation), in millimeters, on bottom. EC: error correction, NLR: non-linear registration, PBS: patch-based segmentation. The two top performing methods are emboldened in each column.}
\label{tab:hc_comparison}
\end{table} 

\begin{figure*}[!ht]
    \centering
    \includegraphics[scale=0.8]{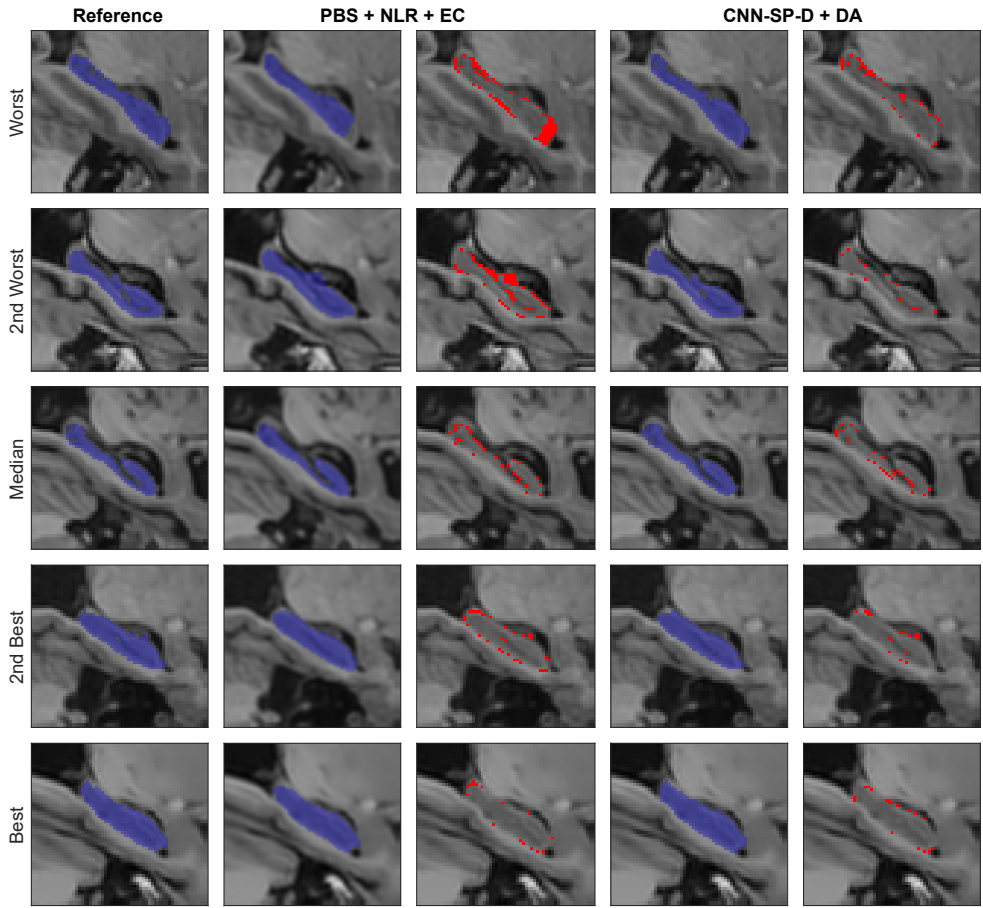}
    \caption{Example right hippocampus segmentations and respective errors using NLR + PB + EC and our best performing method CNN-SP-D + DA. The subjects with the worst, second worst, median, second best and best overlaps after applying NLR + PB + EC are shown for comparison. Errors are overlaid in red in columns three and five.}
    \label{fig:hc_comparison_segs}
\end{figure*}

Compared to the best method from \citep{Zandifar2017}, which combines patch-based segmentation with non-linear registration and error correction (PBS + NLR + EC), our best performing method (CNN-SP-D + DA) yielded an improvement of 2.1\% in terms of mean Dice and a decrease in mean MHD of 0.17 mm (over both left and right hippocampi), both of which were statistically significant ($p \leq 10^{-9}$). Example hippocampal segmentations comparing PBS + NLR + EC to CNN-SP-D + DA are displayed in Figure \ref{fig:hc_comparison_segs}. We note that our method better avoids many of the errors made by the multi-atlas segmentation method, even after applying error correction to the latter.

\subsection{Applications to sub-cortical segmentation in IBSR dataset and multi-structure segmentation in OASIS scan-rescan dataset} 

To demonstrate the versatility of our method, we further applied it to sub-cortical segmentation in the IBSR dataset, and multi-structure segmentation in the OASIS scan-rescan dataset. Dataset details are provided in the respective sections below.  We note that for the subsequent experiments, the network architecture is almost identical to that of the full proposed method used in the hippocampus segmentation experiments - only the number of output channels $C$ (e.g. the number of classes) was changed for each respective segmentation task.

\subsubsection{Sub-cortical segmentation in the IBSR dataset}
The IBSR dataset consists of 18 T1w scans which have been previously manually segmented into 32 structures. Of the 32 structures, as done in Dolz et al. (\citeyear{Dolz2017}), we considered the left and right thalamus, caudate, putamen and pallidum, for a total of 9 classes (one class being background). Though the IBSR images are already roughly aligned, they differ in voxel sizes (ranging from 0.84 $\times$ 0.84 $\times$ 1.5 mm$^3$ to 1 $\times$ 1 $\times$ 1.5 mm$^3$) and would likely benefit from a finer-grained registration. However, to demonstrate the robustness of our approach to small misalignments, we opted against this refinement step and used the images without additional pre-processing. To obtain a segmentation for each subject, we carried out a 6-fold cross-validation. We compared several variants of our method to the methods of Dolz et al. (\citeyear{Dolz2017}) and to a 2.5D CNN method (Kushibar et al. (\citeyear{Kushibar2018})) that uses non-linear registration to incorporate spatial probability maps. We note that for the method of Dolz et al., we used publicly available automated segmentations (https://github.com/josedolz/3D-F-CNN-BrainStruct/tree/master/Results/IBSR) to calculate performance measures and statistical significance tests, whereas for the 2.5D CNN method we include results exactly as reported in Kushibar et al., but could not perform statistical significance tests since the automated segmentations are not available for download. Lastly, we also include results from our application of FIRST \citep{Patenaude2011} from the FMRIB Software Library (FSL) \citep{Jenkinson2012} toolbox.

Table 6 summarizes the performance of the various methods for each sub-cortical structure with respect to mean Dice and mean MHD respectively. Of our CNN-based methods, the performance of the baseline network CNN-B was poorest overall (mean Dice = 86.5\% over all 8 structures). Incorporating spatial priors, CNN-SP produced a large ($p < 1 \times 10^{-9}$) increase in overall performance (mean Dice = 88.5\%). Further deepening the network (CNN-SP-D) produced no significant increase in performance with respect to overall mean Dice ($p = 0.76$), which could be likely attributed to an increased capacity for over-fitting due to the higher modelling capacity of the deeper network combined with the highly limited training data (i.e. only 15 subjects per cross-validation fold) associated with the IBSR dataset. Indeed, regularizing the deeper network using data augmentation (CNN-SP-D + DA) produced a large ($p < 1 \times 10^{-9}$) increase in overlap (mean Dice = 89.5\% over all structures) compared to CNN-SP-D. A similar pattern was observed with respect to mean MHD.

Comparing our methods to those from other works, our best performing method without data augmentation (CNN-SP-D) performed similarly to the 2.5D CNN approach, achieving segmentations with slightly better overlap in the putamen and pallidum, and slightly worse overlap in the thalamus and caudate. However, we emphasize that because the proposed method does not depend on expensive non-linear registration, the mean runtime of CNN-SP-D ((16.4 $\pm$ 2.0) seconds per subject) is much lower than the runtime of approximately five minutes per subject as reported in Kushibar et al. The method of Dolz et al. performed better than our baseline method CNN-B with respect to both mean Dice ($p = 3 \times 10^{-4}$) and mean MHD ($p = 2 \times 10^{-4}$), but was outperformed by each of CNN-SP, CNN-SP-D. Finally, our best performing method using data augmentation, CNN-SP-D + DA, performed best out of all six CNN-based methods. However, we note that the data augmentation scheme used in this work is general and could be used to also boost the performance of the other CNN-based methods under comparison. 

\begin{figure*}[!ht]
    \centering
    \includegraphics[scale=0.8]{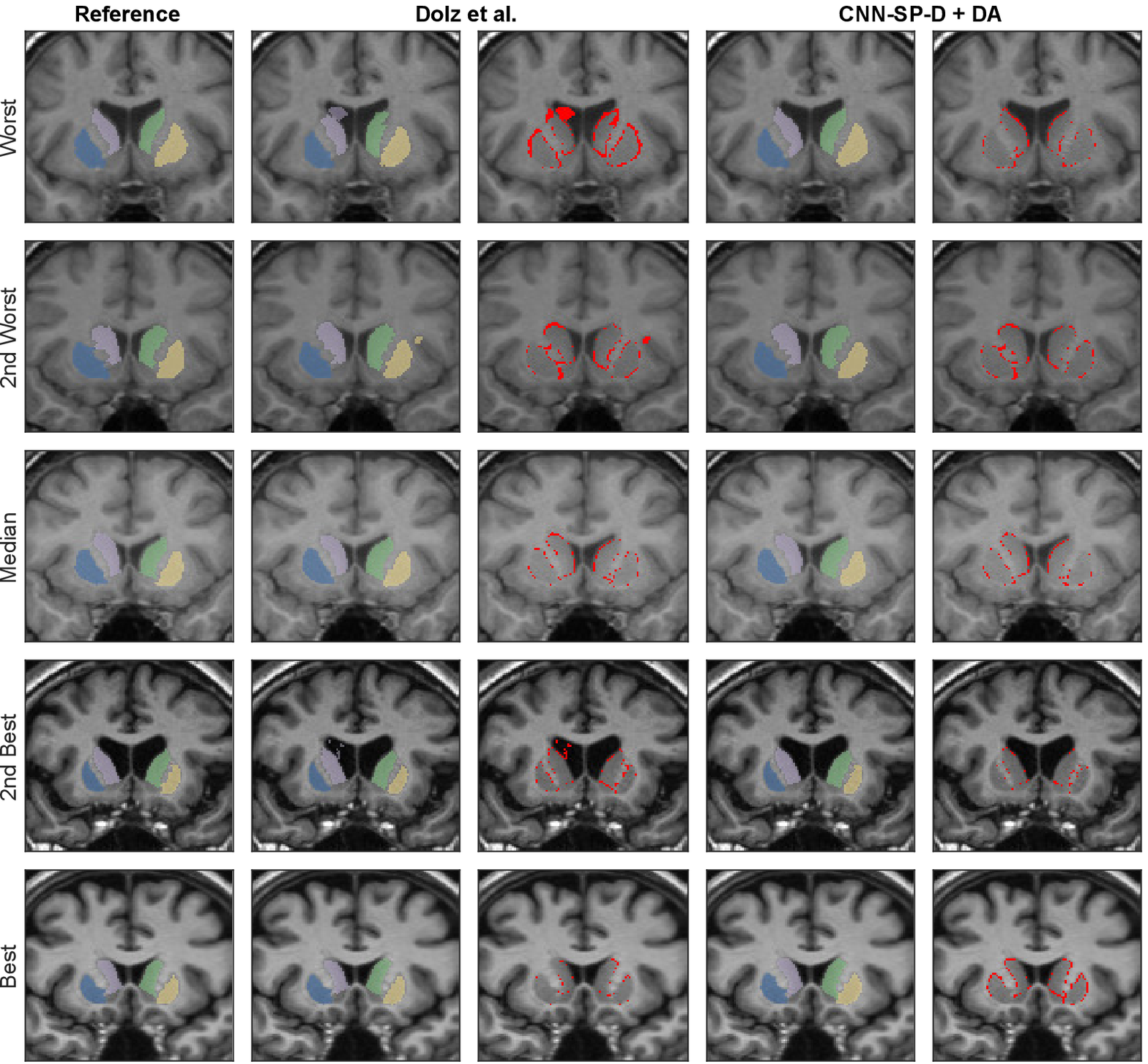}
    \caption{Example sub-cortical segmentations and respective errors using the method of Dolz et al. (\citeyear{Dolz2017}) and our best performing method CNN-SP-D + DA. The subjects with the worst, second worst, median, second best and best overlaps after applying the method of Dolz et al. are shown for comparison. The putamen is overlaid in yellow and blue, and the caudate in light purple and green. Errors are overlaid in red in columns three and five.}
    \label{fig:ibsr_segmentations}
\end{figure*}

Example segmentations comparing the approach of Dolz et al. to our best performing method (CNN-SP-D + DA) are shown in Figure \ref{fig:ibsr_segmentations}. Compared to the former method, our method produced generally smoother segmentations and better avoided more drastic errors typified in the first and second rows of Figure \ref{fig:ibsr_segmentations}. 

\begin{table*}[!ht]
\fontsize{9}{10}\selectfont
\begin{tabular*}{\linewidth}{l @{\extracolsep{\fill}} lllllll }
\hline
& 2.5D CNN & CNN-B & CNN-SP & CNN-SP-D & CNN-SP-D + DA & Dolz et al. & FIRST  \\
\hline
L Thalamus & \textbf{91.0 (1.4)} & 88.3 (2.2) & 90.8 (1.4) & 90.2 (1.4) & \textbf{91.1 (1.2)} & 90.1 (3.1) & 89.9 (1.1)  \\
 & N/A & 0.56 (0.13) & \textbf{0.42 (0.09)} & 0.47 (0.10) & \textbf{0.42 (0.07)} & 0.45 (0.17) & 0.52 (0.07) \\
R Thalamus & \textbf{91.4 (1.6)} & 89.6 (1.9) & 91.0 (1.4) & 90.9 (1.5) & \textbf{91.5 (1.3)} & 90.7 (2.8) & 89.0 (1.3)  \\
 & N/A & 0.58 (0.39) & \textbf{0.43 (0.08)} & 0.43 (0.08) & \textbf{0.41 (0.07)} & 0.44 (0.16) & 0.55 (0.08)  \\
L Caudate & \textbf{89.6 (1.8)} & 86.2 (5.1) & 88.7 (3.8) & 89.2 (2.4) & \textbf{89.9 (2.2)} & 87.7 (6.4) & 82.9 (2.6)  \\
 & N/A & 0.37 (0.16) & 0.28 (0.11) & \textbf{0.27 (0.07)} & \textbf{0.25 (0.06)} & 0.39 (0.46) & 0.42 (0.07)  \\
R Caudate & \textbf{89.6 (2.0)} & 87.4 (5.4) & 88.9 (3.0) & 88.4 (2.6) & \textbf{90.0 (2.7)} & 87.7 (7.3) & 85.0 (4.7) \\
 & N/A & 0.35 (0.25) & \textbf{0.29 (0.10)} & 0.31 (0.09) & \textbf{0.26 (0.10)} & 0.33 (0.29) & 0.35 (0.13) \\
L Putamen & 90.0 (1.4) & 88.5 (2.7) & 90.3 (1.5) & \textbf{90.4 (1.4)} & \textbf{91.0 (1.2)} & 89.0 (4.5) & 88.7 (1.4)  \\
 & N/A & 0.37 (0.12) & \textbf{0.32 (0.08)} & 0.32 (0.09) & \textbf{0.30 (0.07)} & 0.38 (0.25) & 0.41 (0.08)  \\
R Putamen & 90.4 (1.2) & 88.5 (3.0) & 90.3 (1.6) & \textbf{90.5 (1.5)} & \textbf{91.6 (1.3)} & 89.3 (5.4) & 88.6 (1.1) \\
 & N/A & 0.37 (0.12) & 0.32 (0.09) & \textbf{0.32 (0.09)} & \textbf{0.27 (0.06)} & 0.39 (0.35) & 0.42 (0.08)  \\
L Pallidum & 82.6 (5.0) & 82.0 (3.9) & 83.9 (2.6) & \textbf{84.5 (2.8)} & \textbf{85.5 (2.2)} & 82.6 (5.7) & 81.3 (3.8)  \\
 & N/A & 0.47 (0.14) & 0.42 (0.12) & \textbf{0.40 (0.12)} & \textbf{0.38 (0.09)} & 0.44 (0.18) & 0.55 (0.14)  \\
R Pallidum & 82.9 (4.6) & 81.4 (6.1) & 84.0 (2.8) & \textbf{84.4 (3.0)} & \textbf{85.5 (2.7)} & 83.1 (6.3) & 81.8 (3.7)  \\
 & N/A & 0.49 (0.21) & 0.41 (0.12) & \textbf{0.40 (0.12)} & \textbf{0.38 (0.12)} & 0.43 (0.17) & 0.55 (0.15)  \\
All & N/A & 86.5 (4.9) & 88.5 (3.6) & \textbf{88.6 (3.2)} & \textbf{89.5 (3.1)} & 87.5 (6.0) & 85.9 (4.3)  \\
 & N/A & 0.45 (0.22) & \textbf{0.36 (0.12)} & 0.37 (0.11) & \textbf{0.33 (0.11)} & 0.41 (0.27) & 0.47 (0.13)  \\
\hline
\end{tabular*}
\caption{Comparison of various segmentation methods for the segmentation of eight sub-cortical structures in the IBSR dataset. Each table cell reports the mean Dice coefficient (standard deviation) as a percentage on top and the mean MHD (standard deviation), in millimeters, on bottom. The two top performing methods are emboldened in each row.}
\label{tab:ibsr_comparison}
\end{table*} 

\subsubsection{Multi-structure segmentation in the OASIS scan-rescan dataset}
The OASIS scan-rescan dataset contains T1w scans of 20 healthy young adult subjects each scanned on two separate occasions within a period of 90 days. Expert manually generated labels of both session images are available by subscription to Neuromorphometrics (http://www.neuromorphometrics.com). To estimate the scan-rescan reliability of the different labelling methods, second-session images were registered (by estimating a 6-parameter rigid transformation) to the corresponding first-session images, and the labels were propagated using the corresponding rigid transformations with nearest-neighbour interpolation. Accuracy was assessed using a 5-fold cross validation on images from the first session only.

We note that the estimated scan-rescan reliability of manual labelling will necessarily be lower than that of the true reliability of manual labelling (i.e. estimated by comparing multiple manual labellings performed on the same image), the latter being traditionally used as an upper-bound estimate for the attainable accuracy of automated segmentation methods. This is because estimating scan-rescan reliability has two additional sources of error: errors in rigid registration, and artefacts produced during interpolation. Therefore, we consider the scan-rescan reliability of manual labellings as a \textit{lower bound estimate} on the true reliability of manual labellings. While the manual labellings for this dataset were carried out in accordance with a strict protocol, no special effort was made to make the boundaries between regions as smooth as possible. In preliminary studies, we noticed that manual reliability estimates appeared artificially low because of these noisy boundaries. We therefore smoothed the manual labels using median filtering with a small $3 \times 3 \times 3$ kernel, which resulted in higher and more reasonable manual reliability estimates.

For comparison, we also consider FIRST segmentations, since FIRST has been previously shown to be highly reliable \citep{Morey2010}. In our comparisons, for compatibility with FIRST segmentations, we considered the left and right thalamus, caudate, putamen, pallidum, hippocampus and amygdala for a total of 13 classes (one class being background). All CNN-based methods used pre-processed data, with pre-processing consisting of N3 non-uniformity correction, affine registration to the MNI-ICBM152 template space with $1 \times 1 \times 1$ mm$^3$ resolution, and linear intensity normalization. To obtain a segmentation for each subject, we performed a 5-fold cross validation experiment. We applied the same trained network to both session images for each subject in the respective test fold. All label estimates were then resampled back to native space, using nearest-neighbour interpolation, with the inverse of the corresponding transform estimated during the pre-processing stage. 

The scan-rescan reliabilities of manual labelling as well as the automated methods under comparison are reported in Table 7. Each of the automated methods under comparison produced more reliable segmentations compared to manual segmentation. In general, FIRST was comparably reliable (mean Dice = 91.7\%, mean MHD = 0.24 mm, over all 12 structures) to CNN-SP, and more reliable compared to CNN-B (mean Dice = 90.7\%, mean MHD = 0.44 mm). Both CNN-SP-D and CNN-SP-D + DA produced the most reliable segmentations (mean Dice $\geq$ 92.2\%, mean MHD $\leq$ 0.21 mm), and were the most consistently reliable methods, producing small standard deviations on distributions of Dice coefficients and MHD values.  

\begin{table*}[!ht]
\centering
\fontsize{9}{10}\selectfont
\begin{tabular*}{\linewidth}{l @{\extracolsep{\fill}} lllllll }
\hline
& CNN-B & CNN-SP & CNN-SP-D & CNN-SP-D + DA & FIRST & Manual  \\
\hline
L Caudate & 92.4 (1.5) & 92.1 (1.8) & \textbf{92.7 (1.6)} & \textbf{92.7 (1.5)} & 90.8 (5.1) & 87.2 (2.4) \\
 & 0.18 (0.03) & 0.19 (0.05) & \textbf{0.18 (0.03)} & \textbf{0.17 (0.03)} & 0.24 (0.14) & 0.33 (0.06) \\
R Caudate & 92.0 (1.3) & 91.9 (1.3) & \textbf{92.5 (0.9)} & \textbf{92.5 (0.9)} & 91.9 (1.0) & 87.5 (2.5) \\
 & 0.22 (0.14) & 0.19 (0.03) & \textbf{0.18 (0.02)} & \textbf{0.18 (0.02)} & 0.20 (0.02) & 0.33 (0.11) \\
L Putamen & 93.4 (1.6) & 93.7 (1.4) & \textbf{94.0 (1.3)} & 93.9 (1.2) & \textbf{94.5 (0.6)} & 88.9 (2.3) \\
 & 0.20 (0.04) & 0.19 (0.04) & \textbf{0.19 (0.04)} & \textbf{0.18 (0.04)} & 0.20 (0.02) & 0.35 (0.09) \\
R Putamen & 93.2 (1.2) & 93.5 (1.4) & \textbf{94.0 (0.8)} & 93.9 (0.9) & \textbf{94.6 (0.5)} & 89.1 (2.1) \\
 & 0.21 (0.04) & 0.20 (0.04) & \textbf{0.19 (0.03)} & \textbf{0.19 (0.03)} & 0.20 (0.02) & 0.35 (0.08) \\
L Thalamus & 94.5 (1.7) & 95.3 (1.0) & 95.4 (0.8) & \textbf{95.5 (0.7)} & \textbf{96.1 (0.7)} & 91.5 (0.8) \\
 & 0.37 (0.59) & 0.21 (0.04) & 0.20 (0.03) & \textbf{0.20 (0.03)} & \textbf{0.20 (0.03)} & 0.41 (0.04) \\
R Thalamus & 94.8 (0.9) & 95.3 (0.9) & 95.5 (0.7) & \textbf{95.5 (0.7)} & \textbf{95.9 (0.6)} & 91.8 (1.0) \\
 & 0.33 (0.32) & 0.21 (0.04) & \textbf{0.20 (0.03)} & \textbf{0.21 (0.03)} & 0.21 (0.03) & 0.40 (0.06) \\
L Hippocampus & 89.2 (2.1) & 90.9 (1.5) & \textbf{91.1 (1.3)} & \textbf{91.5 (1.2)} & 90.8 (1.3) & 86.6 (1.1) \\
 & 0.24 (0.05) & 0.22 (0.04) & \textbf{0.21 (0.03)} & \textbf{0.20 (0.02)} & 0.24 (0.05) & 0.34 (0.04) \\
R Hippocampus & 89.0 (3.4) & 90.1 (1.8) & 91.0 (1.2) & \textbf{91.3 (0.9)} & \textbf{91.1 (0.8)} & 86.2 (1.1) \\
 & 0.37 (0.61) & 0.24 (0.07) & \textbf{0.21 (0.03)} & \textbf{0.20 (0.02)} & 0.22 (0.02) & 0.36 (0.06) \\
L Pallidum  & 91.3 (2.3) & 92.1 (1.7) & 92.2 (1.8) & \textbf{92.8 (1.4)} & \textbf{92.3 (1.9)} & 85.5 (4.8) \\
 & 0.23 (0.05) & \textbf{0.21 (0.05)} & 0.21 (0.05) & \textbf{0.19 (0.04)} & 0.22 (0.05) & 0.43 (0.14) \\
R Pallidum & 90.5 (1.7) & 91.7 (2.1) & 92.1 (1.0) & \textbf{92.4 (1.3)} & 91.2 (3.0) & 85.7 (3.7) \\
 & 0.24 (0.04) & 0.22 (0.07) & \textbf{0.20 (0.03)} & \textbf{0.20 (0.04)} & 0.25 (0.10) & 0.41 (0.12) \\
L Amygdala & 82.0 (2.0) & 87.7 (2.8) & \textbf{88.1 (2.9)} & \textbf{88.6 (2.7)} & 84.8 (7.3) & 80.8 (4.6) \\
 & 2.37 (8.90) & \textbf{0.26 (0.05)} & 0.26 (0.06) & \textbf{0.24 (0.05)} & 0.34 (0.15) & 0.47 (0.16) \\
R Amygdala & 86.5 (3.1) & 86.3 (8.1) & \textbf{87.6 (2.9)} & \textbf{88.7 (2.5)} & 85.8 (5.4) & 81.4 (2.5) \\
 & 0.28 (0.07) & 0.31 (0.22) & \textbf{0.27 (0.07)} & \textbf{0.25 (0.05)} & 0.32 (0.14) & 0.44 (0.08) \\
All & 90.7 (6.8) & 91.7 (3.8) & \textbf{92.2 (2.9)} & \textbf{92.4 (2.6)} & 91.7 (4.7) & 86.9 (4.2) \\
 & 0.44 (2.65) & 0.22 (0.09) & \textbf{0.21 (0.05)} & \textbf{0.20 (0.04)} & 0.24 (0.10) & 0.39 (0.11) \\
\hline
\end{tabular*}
\caption{Reliability in the OASIS scan-rescan dataset. Each table cell reports the mean Dice coefficient (standard deviation) as a percentage on top and the mean MHD (standard deviation), in millimeters, on bottom. The two top performing methods are emboldened in each row.}
\end{table*}

The accuracy of the automated methods under comparison are reported in Table 8. Despite FIRST being a highly reliable method, it was overall the least accurate method under comparison (mean Dice = 78.0\%, mean MHD = 0.77 mm, over all 12 structures). This is likely due to the fact that FIRST does not learn from user-specified training data, but instead incorporates priors derived from its own training data, which may differ with respect to the anatomical protocol used for manual labelling, as well as with respect to the quality of such manual labels. Example segmentations compared FIRST to our best performing CNN-based method are shown in Figure 7.

Among the CNN-based methods, CNN-B (mean Dice = 84.6\%, mean MHD = 0.50 mm) again performed poorest overall, followed by CNN-SP (mean Dice = 85.0\%, mean MHD = 0.48 mm). It is worth noting that in this dataset, augmenting the baseline network with spatial priors provided a relatively minor performance gain compared to the previous segmentation tasks. Using a deeper network (CNN-SP-D), however, proved particularly beneficial for this task (mean Dice = 86.0\%, mean MHD = 0.42 mm). Finally, CNN-SP-D was further improved using data augmentation (CNN-SP-D + DA), resulting in the best performance (mean Dice = 86.6\%, mean MHD = 0.41 mm). Indeed, a Wilcoxon signed-rank test between between the accuracy of CNN-SP-D + DA and the reliability of manual labellings (mean Dice = 89.9\%, mean MHD = 0.39 mm) showed no significant difference with respect to mean Dice ($p = 0.72$); however, the mean MHD of manual labellings was slightly but significantly lower ($p = 0.02$) compared to CNN-SP-D + DA.

\begin{table*}[!ht]
\centering
\fontsize{9}{10}\selectfont
\begin{tabular*}{\linewidth}{l @{\extracolsep{\fill}} llllll }
\hline
& CNN-B & CNN-SP & CNN-SP-D & CNN-SP-D + DA & FIRST \\
\hline
L Caudate & 85.7 (3.9) & 85.4 (3.8) & \textbf{86.5 (3.1)} & \textbf{87.1 (2.9)} & 71.3 (6.5) \\
 & 0.45 (0.25) & 0.44 (0.16) & \textbf{0.36 (0.09)} & \textbf{0.37 (0.11)} & 0.94 (0.22) \\
R Caudate & 86.5 (3.3) & 85.8 (4.1) & \textbf{87.1 (3.1)} & \textbf{87.1 (2.7)} & 68.4 (7.4) \\
 & 0.42 (0.22) & 0.42 (0.33) & \textbf{0.34 (0.09)} & \textbf{0.34 (0.08)} & 1.00 (0.25) \\
L Putamen & 88.1 (2.3) & 88.5 (3.1) & \textbf{89.2 (3.1)} & \textbf{89.3 (2.7)} & 84.7 (2.1) \\
 & 0.41 (0.10) & 0.37 (0.11) & \textbf{0.35 (0.11)} & \textbf{0.34 (0.10)} & 0.60 (0.09) \\
R Putamen & 88.4 (1.9) & 88.6 (1.9) & \textbf{89.3 (2.2)} & \textbf{89.4 (2.1)} & 84.2 (2.5) \\
 & 0.38 (0.07) & 0.38 (0.07) & \textbf{0.36 (0.09)} & \textbf{0.34 (0.07)} & 0.64 (0.11) \\
L Thalamus & 89.5 (1.6) & 89.5 (2.2) & \textbf{90.4 (2.1)} & \textbf{90.5 (1.7)} & 86.1 (2.3) \\
 & 0.55 (0.09) & 0.55 (0.17) & \textbf{0.48 (0.13)} & \textbf{0.47 (0.10)} & 0.79 (0.13) \\
R Thalamus & 90.5 (1.6) & 90.8 (2.0) & \textbf{91.8 (1.3)} & \textbf{91.6 (1.2)} & 88.0 (1.4) \\
 & 0.65 (0.57) & 0.49 (0.24) & \textbf{0.41 (0.07)} & \textbf{0.43 (0.07)} & 0.68 (0.08) \\
L Hippocampus & 81.4 (0.6) & 84.3 (3.3) & \textbf{85.2 (2.1)} &\textbf{86.9 (1.6)} & 80.0 (1.8) \\
 & 0.52 (0.22) & 0.51 (0.35) & \textbf{0.40 (0.09)} & \textbf{0.34 (0.06)} & 0.60 (0.07) \\
R Hippocampus & 83.1 (0.4) & 84.0 (3.9) & \textbf{85.7 (2.1)} & \textbf{86.2 (1.7)} & 79.7 (2.2) \\
 & 0.44 (0.13) & 0.55 (0.59) & \textbf{0.39 (0.10)} & \textbf{0.39 (0.09)} & 0.60 (0.07) \\
L Pallidum & 83.7 (3.4) & 83.7 (4.5) & \textbf{83.7 (4.6)} & \textbf{84.7 (4.5)} & 77.3 (6.4) \\
 & \textbf{0.50 (0.12)} & 0.51 (0.16) & 0.51 (0.16) & \textbf{0.47 (0.16)} & 0.78 (0.26) \\
R Pallidum & 82.8 (4.5) & 84.7 (4.4) & \textbf{84.9 (4.2)} & \textbf{86.0 (3.9)} & 76.3 (6.7) \\
 & 0.50 (0.15) & 0.46 (0.16) & \textbf{0.45 (0.14)} & \textbf{0.42 (0.13)} & 0.79 (0.24) \\
L Amygdala & 76.6 (5.3) & \textbf{77.6 (5.1)} & 77.2 (5.6) & \textbf{79.1 (5.6)} & 68.8 (9.5) \\
 & 0.62 (0.32) & \textbf{0.54 (0.13)} & 0.58 (0.17) & \textbf{0.53 (0.17)} & 0.98 (0.37) \\
R Amygdala & 78.8 (4.2) & 77.6 (7.6) & \textbf{80.6 (3.0)} & \textbf{81.7 (3.7)} & 70.6 (5.2) \\
 & 0.50 (0.15) & 0.56 (0.24) & \textbf{0.47 (0.09)} & \textbf{0.45 (0.11)} & 0.85 (0.16) \\
All & 84.6 (5.5) & 85.0 (5.7) & \textbf{86.0 (5.1)} & \textbf{86.6 (4.6)} & 78.0 (8.4) \\
 & 0.50 (0.25) & 0.48 (0.27) & \textbf{0.42 (0.13)} & \textbf{0.41 (0.13)} & 0.77 (0.24) \\
\hline
\end{tabular*}
\caption{Accuracy in the OASIS scan-rescan dataset. Each table cell reports the mean Dice coefficient (standard deviation) as a percentage on top and the mean MHD (standard deviation), in millimeters, on bottom. The two top performing methods are emboldened in each row.}
\end{table*}

\begin{figure*}[!ht]
    \centering
    \includegraphics[scale=0.8]{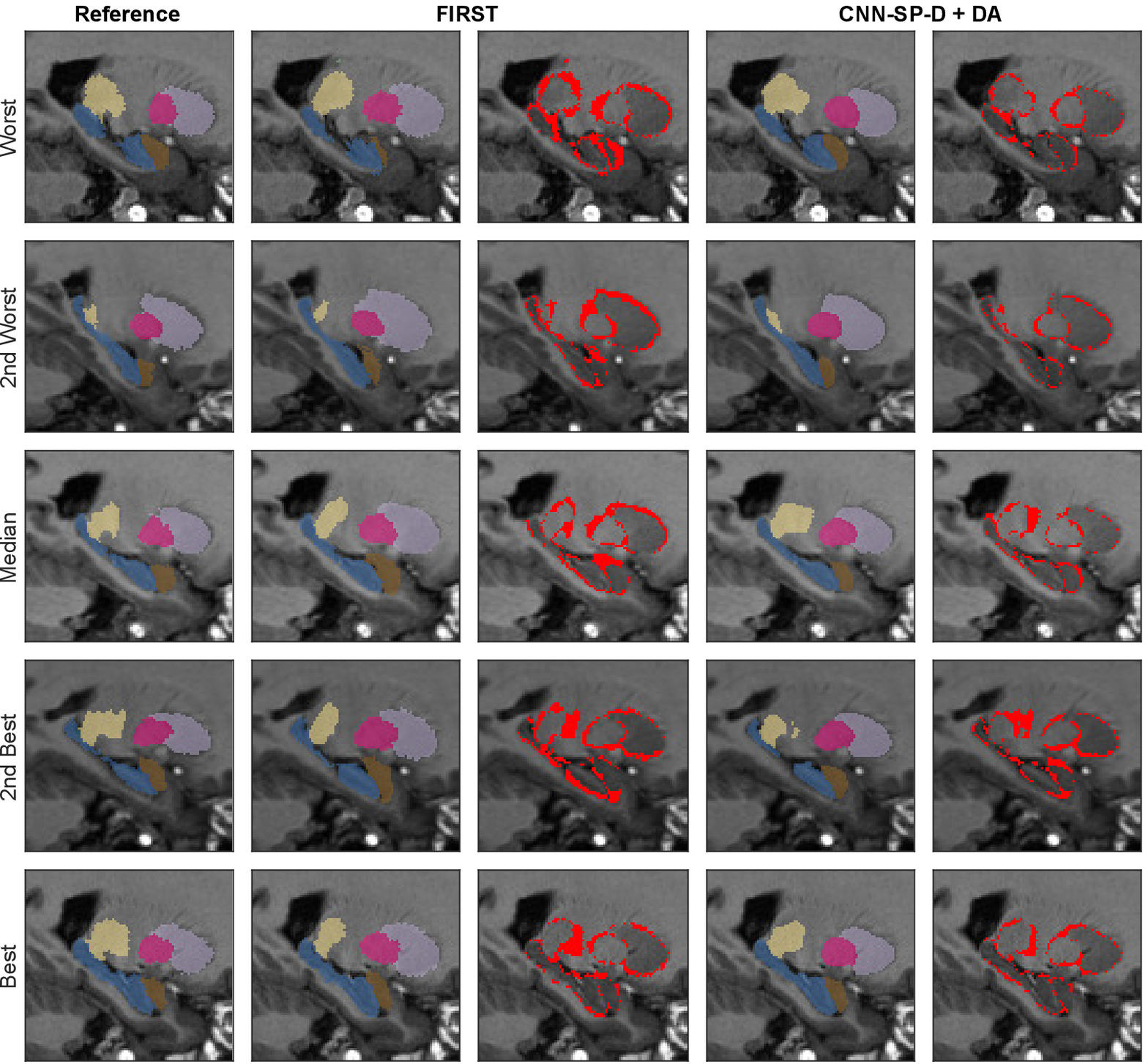}
    \caption{Example multi-structure segmentations and respective errors using FIRST and our best performing method CNN-SP-D + DA. The subjects with the worst, second worst, median, second best and best overlaps after applying FIRST are shown for comparison. The hippocampus is overlaid in blue, amygdala in brown, thalamus in yellow, putamen in pink, and caudate in light purple. Errors are overlaid in red in columns three and five.}
    \label{fig:oasis_segmentations}
\end{figure*}

\section{Discussion}
While many of the other segmentation methods compared in this work also performed reasonably well in general, using a segmentation method with higher accuracy and consistency (i.e. fewer outliers) is always preferable as it increases the separability of populations, enabling controlled trials to be executed with fewer subjects, in turn reducing both cost and duration \citep{Cover2016,Annis2017}. Furthermore, high scan-rescan reliability is a particularly important quality when assessing longitudinal differences in both individuals and groups, and good runtime performance is essential for both clinical and large-scale applications. Finally, unlike many other segmentation methods compared in this work, our proposed method has been demonstrated to be highly versatile and can be easily extended to any segmentation task for which labelled training data are available. In each of three different segmentation experiments, we found our method to be more accurate, robust (marked by fewer outliers) and consistent (marked by lower standard deviations on distributions of Dice coefficients) compared to other state-of-the-art algorithms. Using a scan-rescan dataset, we further demonstrated that the proposed method is highly reliable and produces segmentations of quality comparable to that of expert manual labelling. 

Because of the high degree of regularity in the location of many neuroanatomical structures when normalized to a common space, spatial context is a powerful tool to exploit in MRI segmentation. In Section 4.1.1 it was demonstrated that using spatial priors to assist CNN-based segmentation not only improves performance, but also drastically reduces the computation time required for applying a trained CNN. While a major advantage of deep-learning methods over traditional multi-atlas segmentation is their reduced reliance on extensive pre-processing, our method only requires linear registration to a common space, which is fast (requiring about 20 seconds using a single CPU core), robust (Dadar et al. (\citeyear{Dadar2018}) reports a failure rate of under 0.5\% associated with the linear registration algorithm applied in this work), and in many cases necessary for subsequent processing steps. Further performance gains could possibly be obtained by using non-linear registration to a common template: the use of non-linear registration would, in ideal circumstances, produce more restrictive working volumes (further reducing processing time when applying a trained CNN), and increase the predictive power of spatial coordinates. However, the use of non-linear registration introduces several practical complications: traditional non-linear registration is extremely computationally expensive relative to the time required to apply a trained CNN, and study-specific templates are often required for robust non-linear registration. Combining our approach with deep-learning approaches for non-linear image registration, which have potential for much better computational efficiency, may be a promising avenue for future work. We emphasize however that any performance gains due to the use of spatial priors are to be expected only in proportion to the spatial regularity of the structure of interest. For example, it would not be helpful to use either a working volume or spatial coordinates for the segmentation of brain tumours, which are highly heterogeneous in shape, size, appearance, and location. 

Deep networks like the ones used in this work have a high modelling capacity and are therefore can be more prone to overfitting, particularly when few training samples are available. Indeed this is commonly the case for tasks such as neuroanatomical segmentation, where generating large quantities of high quality training data is a very tedious and time consuming task. While sub-sampling a volume into smaller sub-volumes (patches) is effectively a form of data augmentation, many of the patches extracted from or nearby a particular structure of a given subject will overlap to a large extent (particularly for small structures) and will therefore be somewhat redundant. Overfitting is therefore still possible (as observed in Section 4.1.3), making more aggressive data augmentation schemes necessary for training networks with good generalizability. While many other techniques have been proposed to deal with limited training data (e.g. fine-tuning networks pre-trained on automatic segmentations as done in Roy et al. (\citeyear{Roy2018})), we demonstrated excellent performance using a data augmentation scheme based on random elastic deformations. In the future, more advanced deformation-based techniques could be investigated, e.g. learning a more limited space of plausible deformations using statistical modelling techniques \citep{Onofrey2015,Hauberg2016}, however these methods often come with additional computational costs which may not be justified given the already excellent performance of our proposed method. 

Further contributing to the good performance of the proposed method in cases of very limited training data is the relatively low number of parameters associated with our networks ($\sim 5 \times 10^5$ parameters in our deep network, limiting the capacity to overfit (for comparison, we note that the original U-Net architecture \citep{Ronneberger2015} has $\sim 2 \times 10^7$ parameters). This is in large part due to our choice of using only 32 learnable convolutional filters per layer, since widening the networks showed no appreciable improvement in performance (see Section 4.1.2). On the other hand, increasing the depth of the network (and correspondingly increasing the size of the input patch) resulted in considerable performance gains, which can be attributed to the increased spatial context available to the network in addition to a much higher modelling capacity. Indeed, it has been demonstrated that making networks deeper, as opposed to wider, is a more parameter-efficient way of increasing the modelling capacity of a network \citep{Eldan2016}. While it is likely that the performance of our network could be further improved by fine-tuning the network architecture for specific segmentation tasks, we opted against such an approach to highlight the versatility of this particular network architecture.

A related problem concerns that of generalization across datasets. Since the learned convolutional layers (particularly deeper into the network \citep{Ghafoorian2017b,Kamnitsas2017b}) are highly tailored to the peculiarities of the training data, it is commonly the case that networks trained on a certain dataset perform poorly when applied to an unseen dataset. Nonetheless, robustness to differences between training and testing images (e.g. due to differences in age, health, scanner type, field strength, and/or acquisition sequence) is a highly desirable quality of any method for MRI segmentation. In each of our experiments, we trained and validated our classifiers on challenging multi-centre datasets. However, still more challenging scenarios are commonly encountered in practice; for example, given the often prohibitively high cost of generating high quality manual labellings, it may be desirable to apply a trained classifier to images do not have adequate representation whatsoever in the training set. Future work will address this problem by leveraging so-called `domain adaptation' methods (e.g. \citep{Ganin2016,Hoffman2016}) to learn networks which are robust to differences between the training and target image domains, further increasing the general applicability of our approach.

\section*{Acknowledgements}
\begin{sloppypar}

This work was supported by grants from the Fonds de recherche Santé (FRSQ) and the Healthy Brains for Healthy Lives (HBHL) initiative (made possible with support from the Canada First Research Excellence Fund (CFREF)). We would also like to acknowledge funding from the Famille Louise and André Charron.

Data used in the preparation of this article were obtained from the Alzheimer's Disease Neuroimaging Initiative (ADNI) database (www.loni.ucla.edu/ADNI). As such, the investigators within the ADNI contributed to the design and implementation of ADNI and/or provided data but did not participate in analysis or writing of this report. The ADNI is funded by the National Institute on Aging and the National Institute of Biomedical Imaging and Bioengineering and through generous contributions from the following: Abbott, AstraZeneca AB, Bayer Schering Pharma AG, Bristol-Myers Squibb, Eisai Global Clinical Development, Elan Corporation, Genentech, GE Healthcare, GlaxoSmithKline, Innogenetics NV, Johnson \& Johnson, Eli Lilly and Co., Medpace, Inc., Merck and Co., Inc., Novartis AG, Pfizer Inc., F. Hoffmann-La Roche, Schering-Plough, Synarc Inc., as well as nonprofit partners, the Alzheimer's Association and Alzheimer's Drug Discovery Foundation, with participation from the U.S. Food and Drug Administration. Private sector contributions to the ADNI are facilitated by the Foundation for the National Institutes of Health (www.fnih.org). The grantee organization is the Northern California Institute for Research and Education, and the study was coordinated by the Alzheimer's Disease Cooperative Study at the University of California, San Diego. ADNI data are disseminated by the Laboratory for Neuro Imaging at the University of California, Los Angeles. 

\end{sloppypar}

\bibliographystyle{apalike}
\bibliography{biblio}

\end{document}